\documentclass[%
 reprint,
superscriptaddress,
showkeys,
 amsmath,amssymb,
 aps,
pre, 
longbibliography,
]{revtex4-2}

\usepackage{graphicx}
\usepackage[caption=false]{subfig} 

\usepackage{dcolumn}
\usepackage{bm}

\newcommand{\ee}{\end{equation}} 
\newcommand{\be}{\begin{equation}}

\usepackage[mathscr]{euscript}  
\usepackage{xcolor}  

\usepackage{tcolorbox}
\usepackage{comment}
\usepackage{float}

\makeatletter
\newsavebox{\@brx}
\newcommand{\llangle}[1][]{\savebox{\@brx}{\(\m@th{#1\langle}\)}%
  \mathopen{\copy\@brx\kern-0.5\wd\@brx\usebox{\@brx}}}
\newcommand{\rrangle}[1][]{\savebox{\@brx}{\(\m@th{#1\rangle}\)}%
  \mathclose{\copy\@brx\kern-0.5\wd\@brx\usebox{\@brx}}}
\makeatother

\begin{document}

\preprint{ApS/123-QED}

\title{Optimal diffusion of chiral active particles \\ with strategic reorientations}

\author{Kristian St\o{}levik Olsen}
\thanks{kristian.olsen@hhu.de}

\affiliation{Institut für Theoretische Physik II - Weiche Materie, Heinrich-Heine-Universität Düsseldorf, D-40225 Düsseldorf, Germany}

\author{Hartmut L\"{o}wen}
\affiliation{Institut für Theoretische Physik II - Weiche Materie, Heinrich-Heine-Universität Düsseldorf, D-40225 Düsseldorf, Germany}


\begin{abstract} %
We investigate the competing effects of simultaneous presence of chirality and generalized tumbles in the dynamics of an active Brownian particle. Chiral active particles perform circular motions that give rise to slow transport at late times. By interrupting these circular trajectories at the right time or by performing a tumble at the correct angle, we show that particles can enhance their diffusion. After deriving exact expressions for the orientational propagator and correlations, we use this to calculate the first two moments of displacement. For the effective diffusion coefficient, we study various optimal tumbling strategies. We show that under optimization of the tumbling rate, the case of symmetrically distributed tumbles always give rise to enhanced diffusion, with an effective diffusion coefficient taking a universal value. Next, two cases are considered in detail, namely directional reversal and tumbles at an arbitrary but fixed angle. We discuss how asymmetric tumbles can enhance diffusion beyond that of symmetric tumbles.  Finally, we discuss a situation where the reorientations are realized dynamically in finite time.
\end{abstract}

\pacs{Valid pACS appear here} 
\maketitle

\section{Introduction}

Over the past couple of decades, active matter has fascinated the non-equilibrium statistical physics and soft matter communities with a wide range of intricate phenomena \cite{marchetti2013hydrodynamics,elgeti2015physics,bechinger2016active}. By breaking time-reversal symmetry on the individual particle level, active systems are able to perform tasks such as self-propulsion by continuously absorbing and dissipating energy from and into the environment \cite{o2022time,fodor2022irreversibility}. When parity symmetry is also broken on the particle scale, chiral motion tends to emerge, with circular trajectories of a fixed handedness characterizing the particle dynamics \cite{van2008dynamics,lowen2016chirality}. 

Chiral active motion is found in a wide array of systems, ranging from bacteria swimming near surfaces \cite{frymier1995three,diluzio2005escherichia,lauga2006swimming,hu2015physical,bianchi2017holographic}, to synthetic self-propelled particles such as colloids or granular particles \cite{kummel2013circular,zhang2020reconfigurable,workamp2018symmetry,scholz2021surfactants}. While their motion is intriguing and displays many non-trivial behaviors both on the single-particle and collective level \cite{sevilla2016diffusion,caprini2018active,caprini2019active,caprini2023chiral,liebchen2017collective,debets2023glassy}, the late-time effective diffusion is strongly suppressed when compared to achiral motion \cite{lowen2016chirality}. Hence, chiral motion is by itself not a good strategy of motion if the goal is to increase the distance traversed in a given time. However, when chirality is present in the dynamics simultaneously with other modes of swimming, competing effects can appear. 

Recently, it has become clear that strategic interruptions of chiral trajectories can strongly benefit the late-time diffusive transport. In recent experiments investigating the motion of \emph{Escherichia Coli} near surfaces, it was seen that intermittent stops at a correct rate allows particles to optimize their diffusion \cite{perez2019bacteria}. Similarly, a reversal in the sign of the chirality at the correct time interrupts the circular swimming trajectories and gives rise to an effective diffusion coefficient with multiple local maxima as the chirality is varied \cite{olsen2021diffusion}.{ \color{black} Furthermore, chirality in the presence of correlated noise can lead to enhanced diffusion \cite{weber2011active,bayati2022memory}.} In light of these findings, it is opportune to investigate strategic ways to abort circular trajectories to enhance spatial transport.

\begin{figure}[t]
    \centering
    \includegraphics[width=8.7cm]{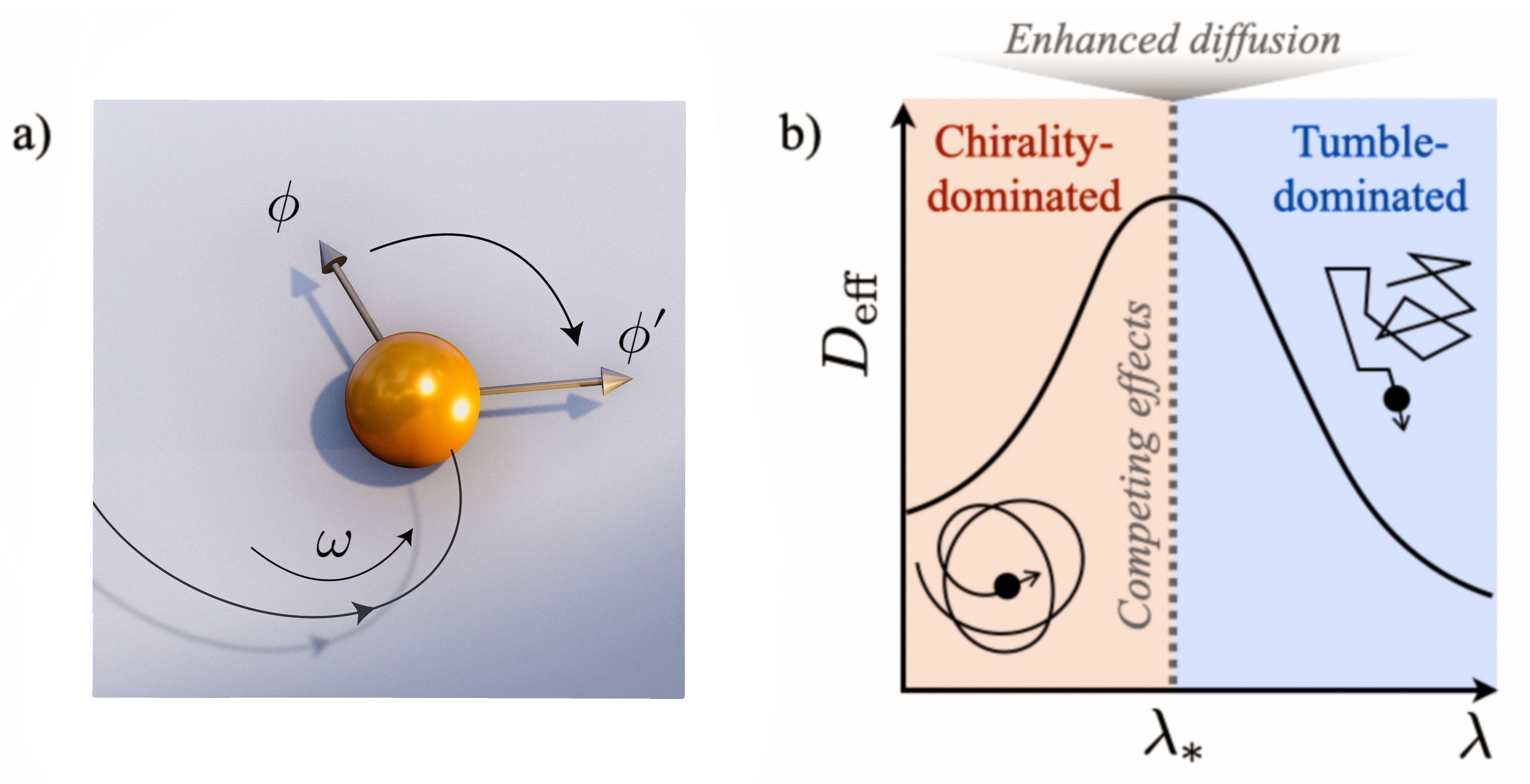}
    \caption{a) Active particles with a chirality $\omega$ perform random reorientations by tumbling an angle $\theta$, drawn from a distribution $b(\theta)$. Tumbles take place at a constant rate $\lambda$. b) When the rate of tumbling is either low ($\lambda \ll \lambda_*$) or high ($\lambda \gg \lambda_*$), transport is suppressed due to circular trajectories or rapid tumbles respectively. For optimal rates $\lambda \approx \lambda_*$ enhanced diffusion takes place. }
    \label{fig:fig1}
\end{figure}

Here, we consider competing effects resulting from the simultaneous presence of both chirality and discrete tumbles in the motion of a self-propelled active Brownian particle (see Fig. \ref{fig:fig1}). We consider tumbling angles that follow any normalized probability density on the circle, symmetric or asymmetric. We fully characterize the directional correlations for the particle, and provide exact expressions for both the first and second moment of displacement. Using these results, we consider several examples of tumbling strategies where the late-time effective diffusion coefficient can be optimized. After deriving a general expression for the optimal tumbling rate that maximizes the effective diffusivity, we consider two cases in detail; directional reversals and asymmetric tumbling strategies. 

{ \color{black} This paper is organized as follows. Section 2 provides the theoretical background of the model we consider, and derives a range of general results, including a general expression for the optimal tumbling rate. Section 3 considers the particular case of directional reversals, and section 4 deals with asymmetric tumbles. In section 5, we consider finite-time reorientations that dynamically realize the tumbling distributions, before a concluding discussion is offered in section 6.}

\section{General theoretical aspects}

Both chiral motion and run-and-tumble dynamics have a long history in statistical physics and fluids dynamics. Submerged bodies with circular trajectories have been studied for decades in the context of asymmetric particles in stationary or sheared liquids \cite{witten2020review}. In the context of active matter, circular paths have been known to take place in the swimming patterns of a wide range of bacteria and other microorganisms, documented for example by H. Berg \cite{berg1990chemotaxis}, but also mentioned in the notes of A. P. van Leeuwenhoek during his studies of \emph{Animalicules} more than three centuries earlier \cite{leeuwenhoek12observations}. From a theoretical perspective, however, chiral self-propelled motion is a relatively recent endeavor \cite{van2008dynamics}. 

Tumbling motion also has a long and rich history. Movement in straight lines interrupted by directional switches in one dimension was a very simple way of generating random motion, and has been studied in various forms over the last hundred years in the context of heat transport or random walks \cite{taylor1922diffusion,goldstein1951diffusion,gupta1958diffusion,kac1974stochastic}. In the early 1970s, many new observations of \emph{Escherichia coli} trajectories emerged, one famous study being that of Berg and Brown in 1972 \cite{berg1972chemotaxis}. A few years later, Lovely and Dahlquist presented a stochastic model of run and tumble motion in both two and three dimensions directly motivated by the recent experiments by Berg and Brown \cite{lovely1975statistical}. In particular, this is the earliest reference (to our knowledge) where transport properties under a general distribution of tumbles angles was considered. Since then, a wide range of generalizations have been considered \cite{schnitzer1993theory,othmer2000diffusion,othmer2002diffusion,erban2004individual,tailleur2008statistical,polin2009chlamydomonas,thiel2012anomalous,malakar2018steady,villa2020run,mori2020universal,frydel2022run, santra2020run,breoni2022one,loewe2024anisotropic}. 

Here, we extend the conventional chiral active Brownian particle model by including tumbles, where the reorientation angle follow a general distribution $b(\phi)$. Tumbles take place at a constant rate $\lambda$. The equations of motion take the form 
 \begin{align}
     \dot {\bm{x}} &= v_0 \hat e(t) =  v_0 [\cos\phi,\sin\phi]\\
     \dot \phi &= \sqrt{2D_r} \xi(t) + \omega +\sum_{\alpha} \theta_\alpha \delta(t-t_\alpha)
 \end{align}
where $\{\theta_\alpha\}$ are random (independent) tumbling angle drawn from a distribution with density $b(\theta)$, and $\{t_\alpha\}$ is a sequence of random instances of time generated by a Poisson process with rate $\lambda$. In the above, $D_r$ is the rotational diffusion coefficient, which determines the strength of the Gaussian white noise $\xi(t)$. The self-propulsion speed is given by $v_0$, and $\omega$ is the chirality. See Fig. \ref{fig:fig1} for a sketch of the dynamics. The associated Fokker-Planck equation can be obtained by the probability balance equation 
\begin{align}
    p(\bm{x},\phi,t+dt) &= \lambda dt \langle p(\bm{x},\phi - \theta,t)\rangle_\theta \\
    &+ (1-\lambda dt)\langle p(\bm{x}-{\Delta_{\bm{x}}},\phi-\Delta_{\phi},t)\rangle_{\Delta_{\bm{x}},\Delta_{\phi}}\nonumber
\end{align}
where $\Delta_{\bm{x}}= v_0 dt \hat{e}(t)$ and $\Delta_\phi = \sqrt{2D_r} dW(dt)$ are infinitesimal steps associated with the stochastic dynamics \emph{without} tumbles $(\lambda = 0)$. The first term on the right-hand-side comes from trajectories where a tumble took place in $(t,t+dt)$, which happens with probability $\lambda dt$. In this case, the contributions to the propagator $p(\bm{x},\phi,t+dt)$  comes from state-space points $(\bm{x},\phi - \theta)$ that tumble into $(\bm{x},\phi)$. Similarly, the second term comes from trajectories with no tumble in $(t,t+dt)$, in which case the dynamics simply evolves according to the stochastic equations of motion with $\lambda=0$.

Expanding in small $dt,\Delta_{\bm{x}},\Delta_{\phi}$ and letting $dt\to 0$ results in the Fokker-Planck equation
\begin{align}
    \partial_t p(\bm{x},\phi,t) &= - v_0 \nabla_{\bm{x}} \cdot \left[ \hat ep(\bm{x},\phi,t) \right] + D_r\partial_\phi^2 p(\bm{x},\phi,t)\nonumber\\
    &- \omega \partial_\phi p(\bm{x},\phi,t) \nonumber\\
    & - \lambda p(\bm{x},\phi,t) + \lambda\int d\theta b(\theta)p(\bm{x},\phi - \theta,t)
\end{align}
Similar composite dynamics with both rotational diffusion and tumble dynamics was studied in the 90s by Schnitzer \cite{schnitzer1993theory}, but to the best of our knowledge the combined effect of chirality and generalized tumbles with a distribution $b(\theta)$ has not been studied.

To build an intuition for the dynamics of the above model before proceeding with a formal analysis, it is instructive to consider an expansion in small tumbling angles $\theta$. We expand to second order in the integrand
\begin{align}\label{eq:exp}
    \int d\theta b(\theta)p(\bm{x},\phi - \theta,t) &\approx p(\bm{x},\phi,t) - \langle \theta\rangle \partial_\phi p(\bm{x},\phi,t) \\
    &+ \frac{\langle \theta^2\rangle}{2} \partial_\phi ^2p(\bm{x},\phi,t)\nonumber
\end{align}
Combining this with the Fokker-Planck equation, we find the effective description
\begin{align}
    \partial_t p(\bm{x},\phi,t) &= - v_0 \nabla_{\bm{x}} \cdot \left[ \hat ep(\bm{x},\phi,t) \right] \nonumber\\
    &+ \left[D_r+\lambda\frac{\langle \theta^2\rangle}{2}\right]\partial_\phi^2 p(\bm{x},\phi,t) \nonumber\\
    &- [\omega + \lambda \langle \theta\rangle ] \partial_\phi p(\bm{x},\phi,t) \label{eq:eff_small}
\end{align}
We see that in the limit of small tumbling angles, the first order correction shifts the chirality, while the second order correction contributes to the angular diffusion. It is also worth noting that in the case of symmetric tumbling distributions, all odd moments vanish, and there is no correction to the chirality. 

From the above, we see that the tumbling chiral ABP in the small tumbling angle limit behaves as a chiral ABP with renormalized parameters. The higher order corrections to Eq. (\ref{eq:exp}) have no analogy in the existing chiral ABP model, and will introduce new terms in the equation.  In the following, we provide an exact (non-perturbative) solution to the angular dynamics, and discuss the generalization of these small-angle results to the general case.

\subsection{Angular propagator and correlation function}
Many interesting observables can be calculated based on the knowledge of angular dynamics and correlations. Integrating the Fokker-Planck equation over space, we have the angular equation of motion
\begin{align}
    \partial_t \rho(\phi,t|\phi_0) &=  D_r\partial_\phi^2 \rho(\phi,t|\phi_0) - \omega \partial_\phi \rho(\phi,t|\phi_0) \nonumber\\
    & - \lambda \rho(\phi,t|\phi_0) + \lambda\int d\theta b(\theta) \rho(\phi - \theta,t|\phi_0).
\end{align}
where $\rho(\phi,t|\phi_0) =\int d\bm{x} p(\bm{x},\phi,t|\bm{x}_0,\phi_0)$ is the marginalized angular propagator. We also used the fact that the angular dynamics is not coupled to the spatial dynamics, implying that $\rho(\phi,t|\bm{x}_0 , \phi_0) = \rho(\phi,t|\phi_0)$. This equation can conveniently be solved by Fourier series. We write
\begin{equation}
    \rho(\phi,t|\phi_0) =\sum_{n=-\infty}^\infty \tilde \rho_n(t|\phi_0) e^{-i n\phi}
\end{equation}
where the Fourier coefficients are given by 
\begin{equation}
     \tilde \rho_n(t|\phi_0) = \frac{1}{2\pi}\int_{-\pi}^\pi d\phi  \rho(\phi,t|\phi_0) e^{i n \phi}
\end{equation}
The Fokker-Planck equation then gives an equation of motion for the coefficients as
\begin{align}
    \partial_t  \tilde \rho_n(t|\phi_0)& =  \left[-D_r n^2 + i \omega n -\lambda+2\pi \lambda \tilde b_n \right] \tilde \rho_n(t|\phi_0)\\
    &= -\mu_n   \tilde \rho_n(t|\phi_0)
\end{align}
where we defined
\begin{equation}
    \mu_n \equiv D_r n^2 - i \omega n + \lambda(1-2 \pi\tilde b_n)
\end{equation}
These numbers $\mu_n$, which may be real or complex, determine the dynamics of the particle's direction of motion. In the above we have also used the Fourier series of the tumbling angle density 
\begin{equation}
    b(\phi) =\sum_{n=-\infty}^\infty \tilde b_n e^{-i n\phi}
\end{equation}
In total, we then have the angular propagator
\begin{equation}
    \rho(\phi,t|\phi_0) = \frac{1}{2\pi}\sum_{n=-\infty}^\infty e^{-\mu_n t} e^{-in(\phi-\phi_0)}
\end{equation}
where we used a Dirac delta function at $\phi=\phi_0$ as initial condition. We observe that $\mu_n$ has the property $\mu_n^* =\mu_{-n}$, which ensures that  $\rho(\phi,t|\phi_0)$ is real. This follows from the fact that the same property holds for the tumbling distribution, i.e. $\tilde{b}^* =\tilde{b}_{-n}$. Additionally, we observe that the angular propagator has a homogeneity property $\rho(\phi,t|\phi_0)=\rho(\phi-\phi_0,t|0)$.

Combining the above, one readily finds the angular correlation function $ \mathcal{C}(t'-t) = \langle \hat{e}(t') \cdot \hat{e}(t)\rangle$ by the integral 
\begin{align}
   \mathcal{C}(t'-t) &= \int d\phi d\phi'   \cos(\phi'-\phi) \rho(\phi',t'|\phi)\rho(\phi,t|\phi_0)\\
     &= \frac{e^{-\mu_1 (t'-t)} + e^{-\mu_{-1} (t'-t)}}{2}
\end{align}
By using the fact that $\mu_{-1} = \mu_1^*$ we can write
\begin{align}
   \mathcal{C}(t'-t)  &= \frac{e^{-\mu_1 (t'-t)} + e^{-\mu_{1}^* (t'-t)}}{2}
\end{align}
Since $\mu_1\in \mathbb{C}$ is in general a complex number, we decompose it into real and imaginary parts, leading to 
\begin{align}\label{eq:corr}
   \mathcal{C}(t'-t)  &= e^{-\Re[\mu_1](t'-t)} \cos\left(\Im[\mu_1](t'-t)\right)
\end{align}
When there are no tumbles $(\lambda=0)$ we have $\Re[\mu_1] = D_r$ and $\Im[\mu_1] = -\omega$, and we recover the standard results for chiral active particles. { 
Based on this analogy, we define an  effective persistence time and an effective chirality that takes into account the effect of tumbles:
\begin{align}
 \tau_\text{eff}^{-1} \equiv   \Re[\mu_1] &= D_r + \lambda - \lambda \int _{-\pi}^\pi d \theta \:b(\theta)\cos(\theta) \label{eq:re}\\
 \omega_\text{eff} \equiv -  \Im[\mu_1] &=  \omega +  \lambda \int _{-\pi}^\pi d \theta \:b(\theta)\sin(\theta)\label{eq:im}
\end{align}

We notice that the effective persistence time, which governs the exponential decay of the correlation function, gets a contribution from the symmetric part of the tumbling density. Likewise, the effective chirality, which governs the oscillatory part of the correlation function, only gets a contribution from tumbles when the tumbling angle density has an antisymmetric component.  Finally, we note as a matter of consistency that to second order in the tumbling angle, these effective parameters coincide with those in Eq. (\ref{eq:eff_small}).  }

It is worth emphasizing that in strong contrast to the conventional chiral ABP case, the effective persistence and chirality are no longer {  always} independent parameters, and competing effects may arise. For example, trying to minimize the effective chirality so that trajectories become maximally rectilinear may also cause a significant decrease in the effective persistence time, making the trajectories noisy and irregular. Even when tumbles are symmetric and $\omega_\text{eff}=\omega$, tumbles can disrupt circular particle trajectories and strongly affect the dynamics. These effects will be important later when we discuss the effective diffusion coefficient of this model under various tumbling strategies.

\subsection{Mean displacement}
\begin{figure}[t]
    \centering
    \includegraphics[width = 8.65cm]{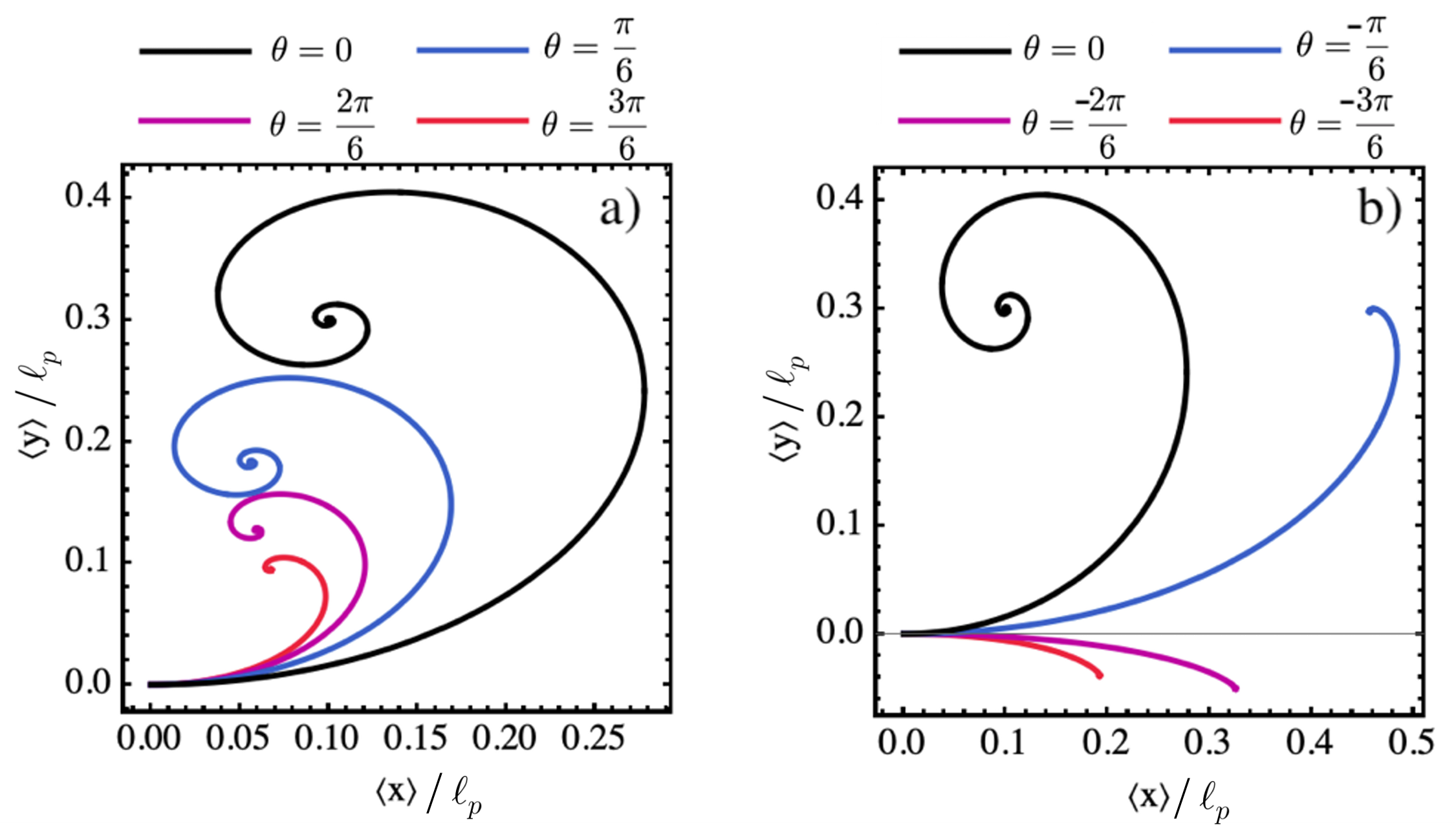}
    \caption{Mean position of the particle under asymmetric tumbles at an angle $\theta$, measured in units of the persistence length $\ell_p = v_0/D_r$. Panel a) shows positive angles, while panel b) shows the corresponding negative tumbling angles. Parameters are set to $D_r = 1, v_0 = 1, \phi_0=0, \lambda = 4, \omega = 3$.}
    \label{fig:spirals}
\end{figure}

Before considering the mean squared displacement and transport properties, we consider the mean displacement. We consider a fixed initial orientation $\phi_0$. We then have
\begin{equation}
    \langle \dot{\bm{x}} \rangle = v_0 \langle \hat{e}(t)\rangle =v_0 \langle e^{i\phi(t)}\rangle 
\end{equation}
where we have identified the 2D vector with a complex number. We have
\begin{equation}
     \langle e^{i\phi(t)}\rangle   = \int_{-\pi}^\pi d\phi e^{i\phi} \rho(\phi-\phi_0,t)
\end{equation}
Changing integration variable to $\vartheta = \phi-\phi_0$, we have 
\begin{equation}
    \langle e^{i\phi(t)} \rangle = e^{i\phi_0}\int_{-\pi}^\pi d\vartheta e^{i \vartheta} \rho(\vartheta,t) = e^{i\phi_0-\mu_1}
\end{equation}
The spatial components of the mean displacement can then be obtained as real and imaginary parts of this complex exponential, which leads to{ 
\begin{equation}
    \langle \dot{\bm{x}}  \rangle = v_0  e^{- t/\tau_\text{eff}}\begin{bmatrix}
\cos(\omega_\text{eff} \: t +\phi_0) \\
\sin(\omega_\text{eff} \: t +\phi_0)
\end{bmatrix}
\end{equation}
}
The mean position can then be obtained simply by integration
\begin{equation}
    \langle {\bm{x}}(t)  \rangle  = \int_0^t dt' \langle \dot{\bm{x}}(t')  \rangle 
\end{equation}
As in the case of a pure chiral ABP, the mean displacement is a \emph{spira mirabilis}, where now the additional tumbling events determine the overall shape of the spiral. Figure \ref{fig:spirals} show some of these spirals for the case $b(\phi) = \delta(\phi-\theta)$.

It is also worth noting that even in the absence of noise $(D_r = 0)$ there can never be a tumbling strategy that results in linear motion, and from the above one can show that $\langle x_i (t)\rangle /t \to 0$ at late times for any tumbling strategy. This is likely different for non-Poissonian tumbling protocols, where one can imagine for example a particle with deterministic directional reversals exactly after a half circle has been spanned, leading to a linear skipping motion.  For Poissonian reorientations, this is never the case.

\subsection{Mean squared displacement}
The mean squared displacement contains much information regarding the dynamics of the system, and is one of the most studied quantities for active single-particle dynamics \cite{bailey2022fitting}. Since we know that the first moment of the displacement is always constant at late times, the mean squared displacement gives the variance of the particle density at late times. Furthermore, if we consider an ensemble where the initial orientation $\phi_0$ is random and uniformly distributed on the circle $[-\pi,\pi]$, $\langle \bm{x} (t) \rangle = 0$ at all times. The mean squared displacement is calculated from the correlations of the orientation vector $\hat e(t)$:
\begin{align}
    \langle \bm{x}^2(t)\rangle &= v_0^2 \int_0^{t}dt'\int_0^{t}dt'' \langle \hat e(t')\cdot \hat e(t'')\rangle \\
    &= v_0^2 \int_0^{t}dt'\int_0^{t}dt'' \mathcal{C}(t',t'') 
\end{align}
where the orientational correlation function is $ \mathcal{C}(t',t'')= \langle \cos[\phi(t')-\phi(t'')]\rangle$. 
From Eqs. (\ref{eq:corr}-\ref{eq:im}) we find the mean squared displacement

\begin{align}\label{eq:msd}
    & \langle\bm{x}^2(t) \rangle = \frac{2v_0^2  \tau_\text{eff}}{1+\tau_\text{eff}^2 \omega_\text{eff}^2} t + \frac{2 \tau_\text{eff}^2 v_0^2 (\tau_\text{eff}^2\omega_\text{eff}^2-1)}{(1 + \tau_\text{eff}^2\omega_\text{eff}^2)^2}\\
     & +\frac{ 2 v_0^2 \tau_\text{eff}^2 e^{-\frac{t}{\tau_\text{eff}}}  \left[ \cos(\omega_\text{eff} t) (\tau_\text{eff}^2\omega_\text{eff}^2-1) + 2 \tau_\text{eff} \omega_\text{eff} \sin(\omega_\text{eff} t) \right]}{(1 + \tau_\text{eff}^2\omega_\text{eff}^2)^2}\nonumber
\end{align}
{ 
When tumbles are symmetric, $\omega_\text{eff} = \omega$ from Eq. (\ref{eq:im}). The mean squared displacement then only depends on the tumbling distribution through $\int d\phi b(\phi) \cos\phi$, which is related to the first Fourier mode of the density $b(\phi)$. This has been shown for achiral tumblers with a symmetric tumbling distribution in the past, e.g., by Taktikos \emph{et al.} \cite{taktikos2013motility} and even earlier in the 1970s by Lovely and Dahlquist \cite{lovely1975statistical}. Since we allow also asymmetric tumbling distributions, an additional dependence on $\int d\phi b(\phi) \sin\phi$ appears in the general case.}

At late times, the effective diffusivity is calculated as
\begin{equation}
    D_\text{eff} \equiv \lim_{t\to \infty} \frac{\langle \bm{x}^2(t) \rangle}{2t}
\end{equation}
By using the above mean squared displacement, Eq. (\ref{eq:msd}), we find
{ 
\begin{align}\label{eq:deff}
    D_\text{eff}(\lambda;[b(\phi)]) = v_0^2 \:\frac{\tau_\text{eff}^{-1}(\lambda;[b(\phi)])}{\tau_\text{eff}^{-2}(\lambda;[b(\phi)])+\omega_\text{eff}^2(\lambda;[b(\phi)])}
\end{align}
Several things are worth noting. First, we note that by construction (through Eqs. (\ref{eq:corr}–\ref{eq:im})), this takes a similar looking form to the normal chiral active Brownian particle case. However, it is worth emphasizing that the dynamics leading to this diffusion coefficient is completely different; the above formula gives the effective diffusion coefficient for any distribution of tumbling angles $b(\theta)$ and tumbling rate $\lambda$. We emphasize the $D_\text{eff}$ depends parametrically on the tumbling rate (as well as on model parameters such as rotational diffusion, self-propulsion speed and chirality) and depends functionally on the tumbling angle distribution $b(\theta)$ through its Fourier coefficients. { \color{black} Secondly, Ref. \cite{perez2019bacteria}  considers a model for intermittent stop-and-go motion of chiral bacteria near surfaces. In this case, after a stop, it was assumed that the particle reorients by an angle distributed with a Heaviside function. In the limit where the stops are instantaneous and the same tumbling distribution is used, we find agreement between Eq. (\ref{eq:deff}) and the results of Ref. \cite{perez2019bacteria}.  Finally,} the special case of uniform tumbles $b(\phi)= (2\pi)^{-1}$ can be recovered by the methods presented in Ref. \cite{Olsen_2024}, where a general framework for predicting effective transport coefficients under symmetric stop-and-go processes was developed. Tumbles in this case correspond to instantaneous stop-and-go events. However, the results presented here are far more general, as we can study optimal reorientation strategies for any $b(\phi)$.
}

\subsection{Optimal tumbling rates}

From the effective diffusion coefficient, Eq. (\ref{eq:deff}), we can derive a general expression for the optimal rate of tumbling, $\lambda_*$, that maximizes the diffusivity for any tumbling dynamics $b(\theta)$. {   As a matter of notation, we recall that the effective persistence time and effective chirality depends on the tumbling rate through

\begin{align}
    \tau_\text{eff}^{-1} &= D_r +\lambda - 2\pi\lambda \Re[\tilde b_1]\\
    \omega_\text{eff} &= \omega + 2\pi\lambda \Im[\tilde b_1]
\end{align}
where we by $\Re[\tilde b_1]$ and $\Im[\tilde b_1]$ denote the real and imaginary parts of the first Fourier coefficient of the tumbling density.  } By differentiating with respect to the rate, we can identify the optimal tumbling rate through $\partial_{\lambda_*} D_\text{eff}(\lambda_*) =0$. After simplifying the algebra, we find 

\begin{equation}\label{eq:optimal_lam}
    \lambda_* = \frac{ \sqrt{\frac{[2\pi D_r\Im[\tilde b_1] + \omega(2\pi\Re[\tilde b_1]-1)]^2}{(2\pi\Im[\tilde b_1])^2 +(2\pi \Re[\tilde b_1]-1)^2}}-D_r}{ 1-2\pi\Re[\tilde b_1]}
\end{equation}
This only exists when the resulting $\lambda_*$ is positive. { \color{black} Intuitively, an optimal tumbling rate should interrupt the circular trajectories at optimal moments in time, resulting in a more persistent trajectory. We see that the optimal tumbling rate is determined not only by the noise $D_r$, but also depends on the specific tumbling dynamics used to interrupt the circular paths, which is encoded in the distribution $b(\theta)$.}

{ 
\subsection{Universal maximum in the case of symmetric tumbles}
In the case of symmetric tumbles, $\Im[\tilde b_1] =0$ and the expression for the optimal tumbling rate simplifies drastically: 
\begin{equation}\label{eq:optimal_lam_sym}
    \lambda_* = \frac{ |\omega| - D_r }{1- 2\pi \Re[\tilde b_1] }.
\end{equation}
We see that a positive solution only exists if $|\omega|> D_r$. The effective diffusion coefficient in this case can through Eq. (\ref{eq:deff}) be shown to take the compact form 
\begin{equation}\label{eq:compact}
    D_\text{eff} = \frac{v_0^2}{2|\omega|}
\end{equation}
We emphasize that this maximum value is universal in the sense that it is independent of the exact tumbling dynamics $b(\theta)$ and the rotational noise. 

For the sake of further intuition regarding this result, we note that the effective persistence time $\tau_\text{eff}$ and effective chirality $\omega_\text{eff}$ are independent in this case, since $\omega_\text{eff} =\omega$ is independent of the tumbling distribution in this case. Hence, the effective diffusivity 
\begin{equation}
D_\text{eff} = \frac{v_0^2 \tau_\text{eff}^{-1}}{\tau_\text{eff}^{-2}+\omega^2}
\end{equation}
can be maximized by finding the optimal value of $\tau_\text{eff}$, independently of what the exact tumbling dynamics is. This optimal effective persistence time is $\tau_\text{eff} = |\omega|^{-1}$, which occurs exactly when the tumbling rate takes the form given in Eq. (\ref{eq:optimal_lam_sym}). Since this statement is independent of the tumbling dynamics, it also holds for the case of no tumbles; in other words, if one could somehow tune the rotational noise strength $D_r$ the maximal diffusion coefficient $D_\text{ch}$ of a chiral active Brownian particle is also given by Eq. (\ref{eq:compact}). The noise giving rise to $D_r$ is externally imposed and often not controllable. Regardless,  this formal argument shows that for any tumbling dynamics, $D_\text{eff}(\lambda_*)\geq D_\text{ch}$ at any fixed value of $D_r$, and the inequality is only attained under the fine-tuning $D_r = |\omega|$. Hence, symmetric tumbles of any type can always be optimized to give rise to enhanced diffusion, and the value of the optimal diffusivity is universally given by Eq. (\ref{eq:compact}).

While it is not generally true that asymmetric tumbles always lead to enhanced diffusion, we will show with en example in section \ref{sec:asym} that for certain tumbling angles diffusion can be enhanced beyond that of Eq. (\ref{eq:compact}).
}

\section{Directional reversals}\label{sec:rev}
\begin{figure}[t]
    \centering
    \includegraphics[width = \columnwidth]{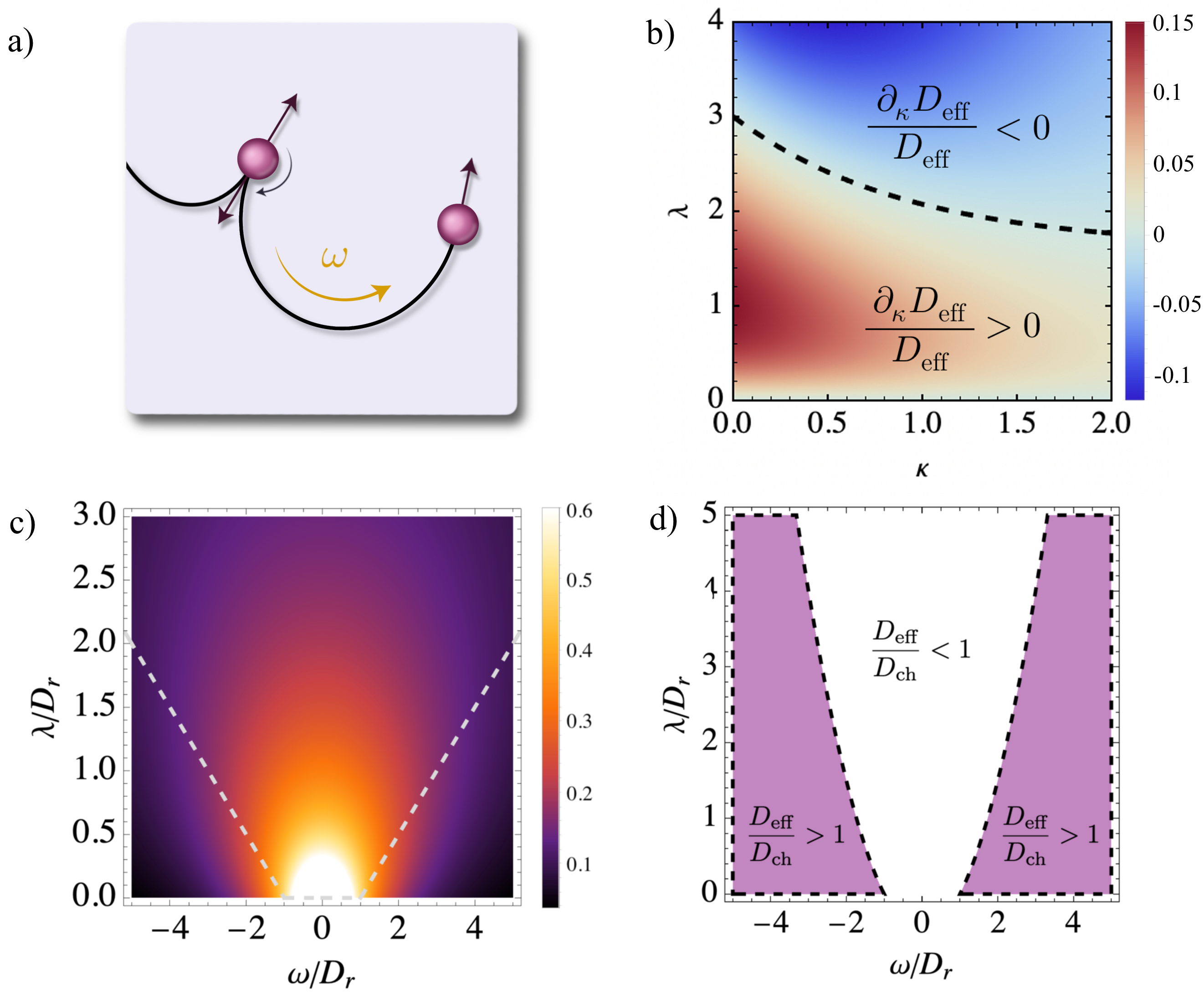}
    \caption{a) An active particle with chirality $\omega$ and directional reversals at random times with rate $\lambda$. b) Slope of the effective diffusivity as a function of reversal rate and reversal accuracy, showing domains of positive and negative slope. c) Effective diffusion coefficient as a function of re-scaled chirality $\omega/D_r$ and reversal rate $\lambda/D_r$. {  The colorbar indicates the value of the effective diffusivity measured relative to the diffusivity $v_0^2/D_r$ of a normal active Brownian particle.} The dashed gray line shows optimal rate $\lambda_*$ for fixed $\omega$. Parameters used are $D_r = 1,v_0 =1$. d) Effective diffusion coefficient measured relative to the case of pure chiral motion $(\lambda=0)$. Shaded areas show parameter space regions where reversals benefit the exploration and gives rise to diffusivity that is greater than that of pure chiral motion. }
    \label{fig:deff_rev}
\end{figure}

The first out of two examples we consider in more detail is the case where the tumbles give rise to reversals in the particle's direction of motion [see Fig. (\ref{fig:deff_rev} a)]. Such behavior is well known to be a part of the swimming patterns of many microorganisms \cite{wu2009periodic,thutupalli2015directional,leonardy2008reversing,taylor1974reversal,johansen2002variability,quelas2016swimming}. The effect of reversals in the absence of chirality has been studied from a theoretical perspective over recent years, including dynamical properties of particles in free space and in confinement \cite{grossmann2016diffusion,villa2020run,santra2021active,santra2021direction}, as well as collective effects \cite{olsen2022collective}. 

If the reversals were perfectly precise, we would have $b(\theta) =\delta(\theta-\pi)$. However, real-world reversals are often imperfect and will not always take place at a turning angle of $180^\circ$. For example, Ref. \cite{johansen2002variability} reports that among a range of marine bacteria that perform reversals, more than 70\% turn at an angle higher than $150^\circ$.  To model such imperfect reversals, we consider a distribution of tumbling angles of the von Mises type
\begin{equation}
    b(\theta) =\frac{1}{2\pi I_0(\kappa)} \exp\left[{\kappa \cos(\theta-\pi)}\right],
\end{equation}
which behaves similarly to a Gaussian on a circle, in this case with a peak at $\theta =\pi$. In the above, $I_n(z)$ is a modified Bessel function of the first kind. The parameter $\kappa$ determines the width of the distribution, with a Dirac delta-peak resulting in the $\kappa \to \infty$ limit. Conversely, when $\kappa\to 0$ conventional tumbling with uniformly distributed tumbling angles will be recovered. We refer to $\kappa$ as the reversal precision.

To use Eq. (\ref{eq:deff}) and Eq. (\ref{eq:optimal_lam_sym}) to find the diffusivity and optimal tumbling angle in the presence of chirality, we need the quantity $\Re[\mu_1]$, while $\Im[\mu_1] =-\omega$ since the tumbling distribution in this case is symmetric. $\Re[\mu_1]$ is in this case given by $\Re[\tilde b_1]$, which we can calculate to be
\begin{align}
    2\pi \Re[\tilde b_1] &= \int_{-\pi}^\pi d\theta \frac{1}{2\pi I_0(\kappa)} \exp\left[{\kappa \cos(\theta-\pi)}\right] \cos\theta\\
    & = -\frac {I_1(\kappa)}{I_0(\kappa)}.
\end{align}
This gives the effective diffusivity and associated optimal tumbling rate
\begin{align}
    D_\text{eff} &= v_0^2\frac{D_r +\lambda+\lambda\frac {I_1(\kappa)}{I_0(\kappa)}}{\left(D_r +\lambda+\lambda\frac {I_1(\kappa)}{I_0(\kappa)}\right)^2+\omega^2}, \label{eq:revdeff}\\
    \lambda_* &= \frac{|\omega|- D_r}{1 +\frac {I_1(\kappa)}{I_0(\kappa)}}.\label{eq:revlam}
\end{align}
It is worth emphasizing that the factor $\frac {I_1(\kappa)}{I_0(\kappa)}$ is a simple monotonically increasing function interpolating between zero and unity as $\kappa$ is increased. The optimal tumbling rate $\lambda_*$ is therefore also monotonic and always decreasing as a function of $\kappa$. This implies that in order to move optimally, the particle has to tumble less frequently if there is a high precision in the reversals. 

{  If one  takes into account various costs associated with sudden reorientations. For example,  it is known that the energetic cost of a tumble depends crucially on the details of the reorientation, e.g., whether the bacteria body has to rotate, or if simply the flagella reverse their rotational direction \cite{mitchell2002energetics}. Alternatively, reorientations take finite time, and this time-overhead can be thought of as a hidden cost associated with each reversal \cite{mori2023optimal}.  The ability to optimize diffusion by performing fewer, but more precise, reversals can then benefit the accumulated cost. However, if higher reversal precision for some reason were to come at a higher cost, a trade-off would appear, and it may be optimal for the particle to sacrifice precision in favor of lower reversal costs. Whatever the situation may be, the optimized diffusion coefficient, taking the universal value given in Eq. (\ref{eq:compact}), is independent of the reversal precision $\kappa$ which can hence be tuned freely without motility losses. }

The effective diffusion coefficient behaves less trivially as the reversal accuracy $\kappa$ is varied when $\lambda \neq \lambda_*$. Figure (\ref{fig:deff_rev} b) shows a plot over parameter space $(\kappa,\lambda)$ of the slope $\partial_\kappa D_\text{eff}(\kappa,\lambda)$, showing clear regions of positive and negative slope. The dashed line shows the $\lambda_*(\kappa)$ line, separating the two regions. For reversal rates $\lambda < \lambda_*(\infty)$ the effective diffusion coefficient always increases monotonically as a function of $\kappa$, while for $\lambda > \lambda_*(0)$ the effective diffusion coefficient decays monotonically as a function of $\kappa$. For intermediate reversal rates $\lambda \in (\lambda_*(\infty),\lambda_*(0))$ there is a non-monotonic behavior as the precision $\kappa$ is varied, which in Figure (\ref{fig:deff_rev} b) is seen from the fact that the dashed line is crossed as $\kappa$ is varied for reversal rates in this range.

Figure (\ref{fig:deff_rev} c) shows the effective diffusion coefficient $ D_\text{eff}$ as a function of (rescaled) reversal rate $\lambda$ and (rescaled) chirality $\omega$ for $\kappa \to \infty$ (e.g. perfect reversal). For fixed $\omega$, the optimal rate $\lambda_* = (|\omega|-D_r)/2$ is shown in a dashed line.  Figure (\ref{fig:deff_rev} d) shows a related phase diagram plot, displaying the regions where the effective diffusion coefficient is greater or lesser than the tumble-free case. We see that, generically, transport can be enhanced by tumbles when chirality is large.

\section{Asymmetric tumbles}\label{sec:asym}

Next, we consider the case where the particle tumbles at an arbitrary angle $\theta$, i.e. $b(\phi)= \delta(\phi-\theta)$. In contrast to directional reversals, these asymmetric tumbles introduce an effective chirality, as has for example been observed in a model of a tumbles that only moves in four principal directions with asymmetric tumbling rates \cite{mallikarjun2023chiral}, and for discrete random walkers that turns at a fixed angle every step \cite{larralde1997transport}. In the present case, we have
\begin{align}
    \tau_\text{eff}^{-1} &=  D_r +\lambda [1-  \cos(\theta)]\\
    \omega_\text{eff} &=  \omega+\lambda \sin(\theta)
\end{align}

\begin{figure}[t]
    \centering
    \includegraphics[width = \columnwidth]{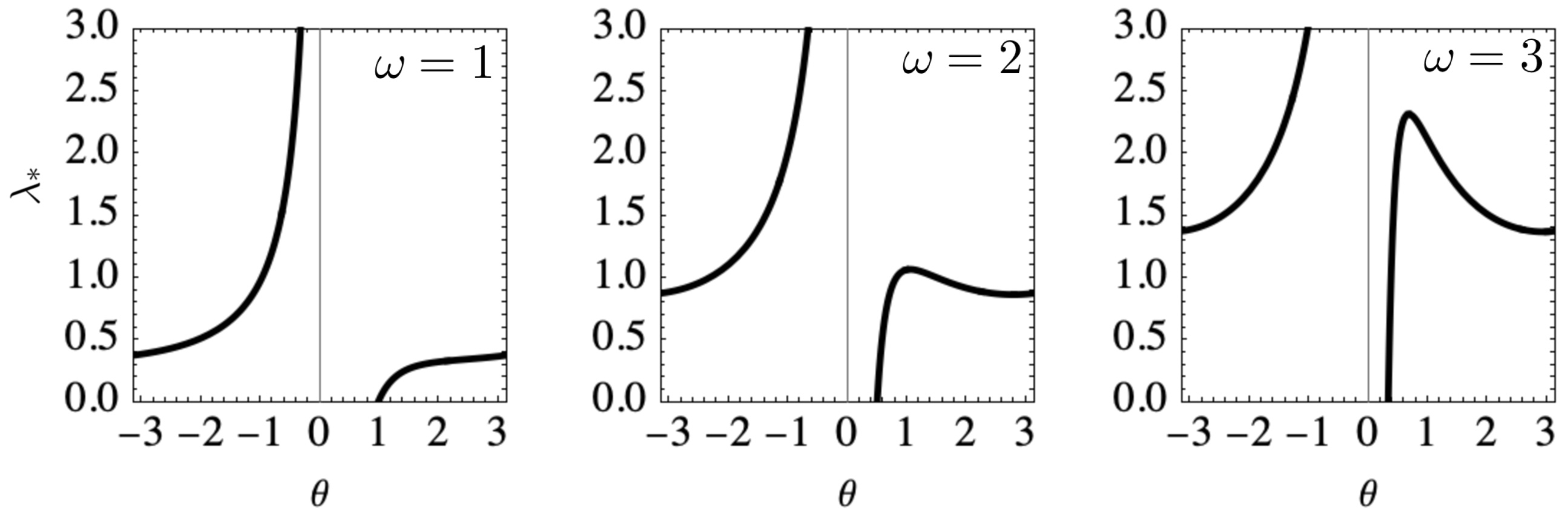}
    \caption{Optimal tumbling rate $\lambda_*$ for a chiral ABP with asymmetric tumbles, as a function of the tumbling angle $\theta$. Panels correspond to different values of the chirality $\omega$ as indicated.  Remaining parameters are set to $D_r = 0.25, v_0 = 1$.}
    \label{fig:asym_lam}
\end{figure}

From Eq. (\ref{eq:deff}) and Eq. (\ref{eq:optimal_lam}), we have the diffusivity and associated optimal rate
\begin{align}
    D_\text{eff} &= v_0^2 \frac{D_r +\lambda -\lambda\cos\theta}{(D_r +\lambda -\lambda\cos\theta)^2+(\omega + \lambda \sin\theta)^2} \\
    \lambda_* &= \frac{D_r}{\cos\theta-1} +\frac{|\omega(\cos\theta-1) + D_r\sin\theta|}{\sqrt{2}[1-\cos\theta]^{3/2}}
\end{align}
We note that the optimal rate does not exist for all choices of $\omega$ and $\theta$, and is considered to exist only when this expression is positive. Figure \ref{fig:asym_lam} shows the optimal tumbling rate as a function of the tumbling angle $\theta$, for different values of the chirality $\omega$. We see that typically there exists a small range of tumbling angles near $\theta=0$ for which there is no optimal tumbling rate. In these cases, the tumbles can never give rise to enhanced diffusion.

\begin{figure}[t]
    \centering
    \includegraphics[width = \columnwidth]{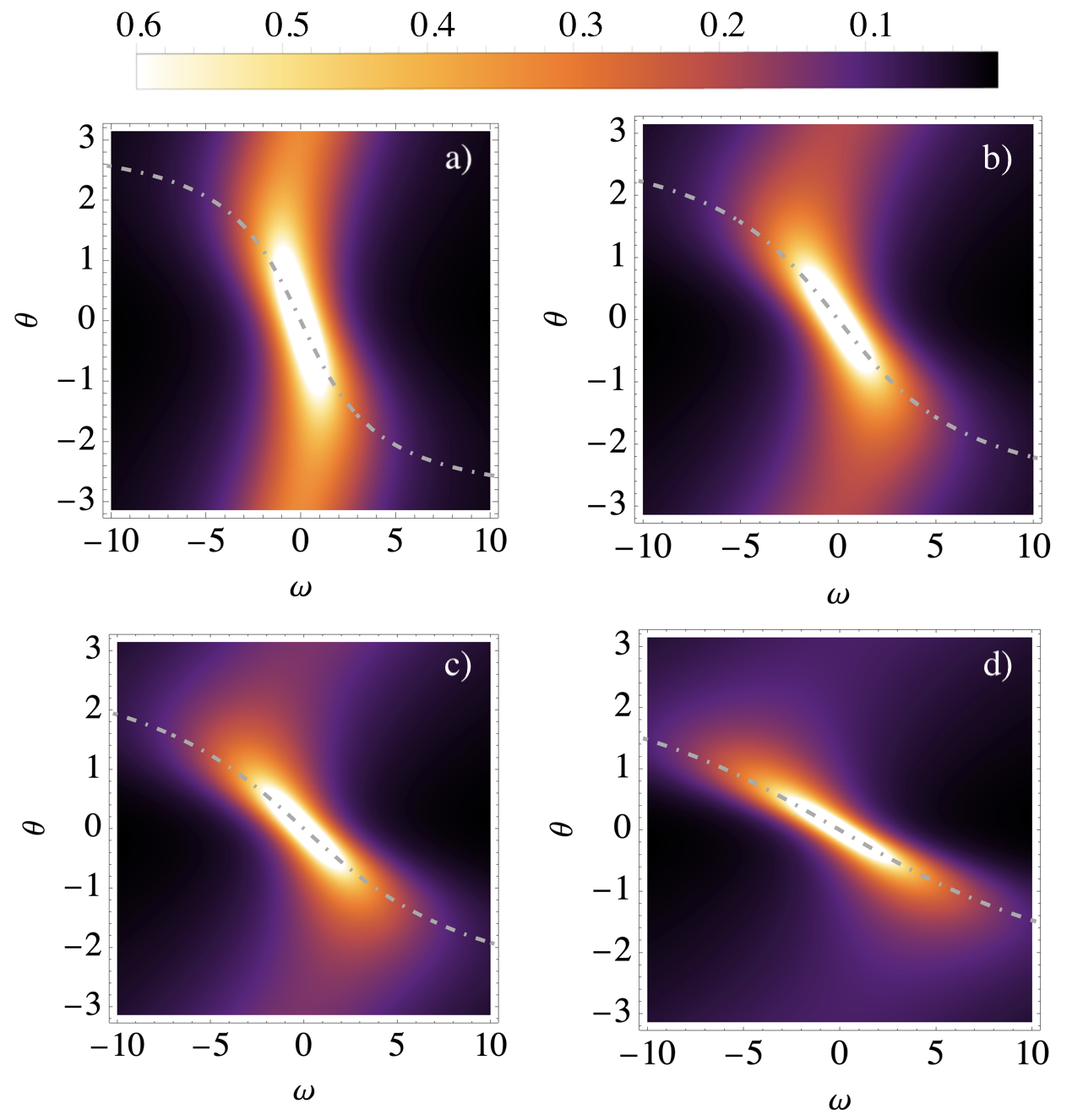}
    \caption{Effective diffusion coefficient as a function of chirality $\omega$ and tumbling angle $\theta$. Panels a)-d) show plots for $\lambda = 1,2,3,5$ respectively. Dashed gray lines show optimal tumbling angle $\theta$ for fixed chirality $\omega$. {  The colorbar indicates the value of the effective diffusivity measured relative to the diffusivity $v_0^2/D_r$ of a normal active Brownian particle.} Parameters are set to $D_r = v_0 = 1$.}
    \label{fig:deff_theta}
\end{figure}

The diffusion coefficient can also be optimized with respect to the tumbling angle $\theta$. Keeping the tumbling rate fixed, we find a maximized diffusivity when
\begin{align}
    \partial_\theta D_\text{eff} \sim&   \left(2 \lambda  D_r+D_r^2-\omega ^2\right)\sin \theta+ 2 \omega   \left(D_r+\lambda \right)\cos \theta  \nonumber \\
    & -2 \lambda  \omega = 0
\end{align}
This can conveniently be turned into a polynomial equation by utilizing trigonometric half-angle identities
\begin{equation}
    \cos \theta = \frac{1-z^2}{1+z^2}\:\:\: ; \:\:\sin \theta = \frac{2z}{1+z^2}
\end{equation}
where $z = \tan(\theta/2)$. Combining the above then yields $z = \omega/(D_r+2\lambda)$. Hence, the optimal tumbling angle is given by 

\begin{figure}[t]
    \centering
    \includegraphics[width = \columnwidth]{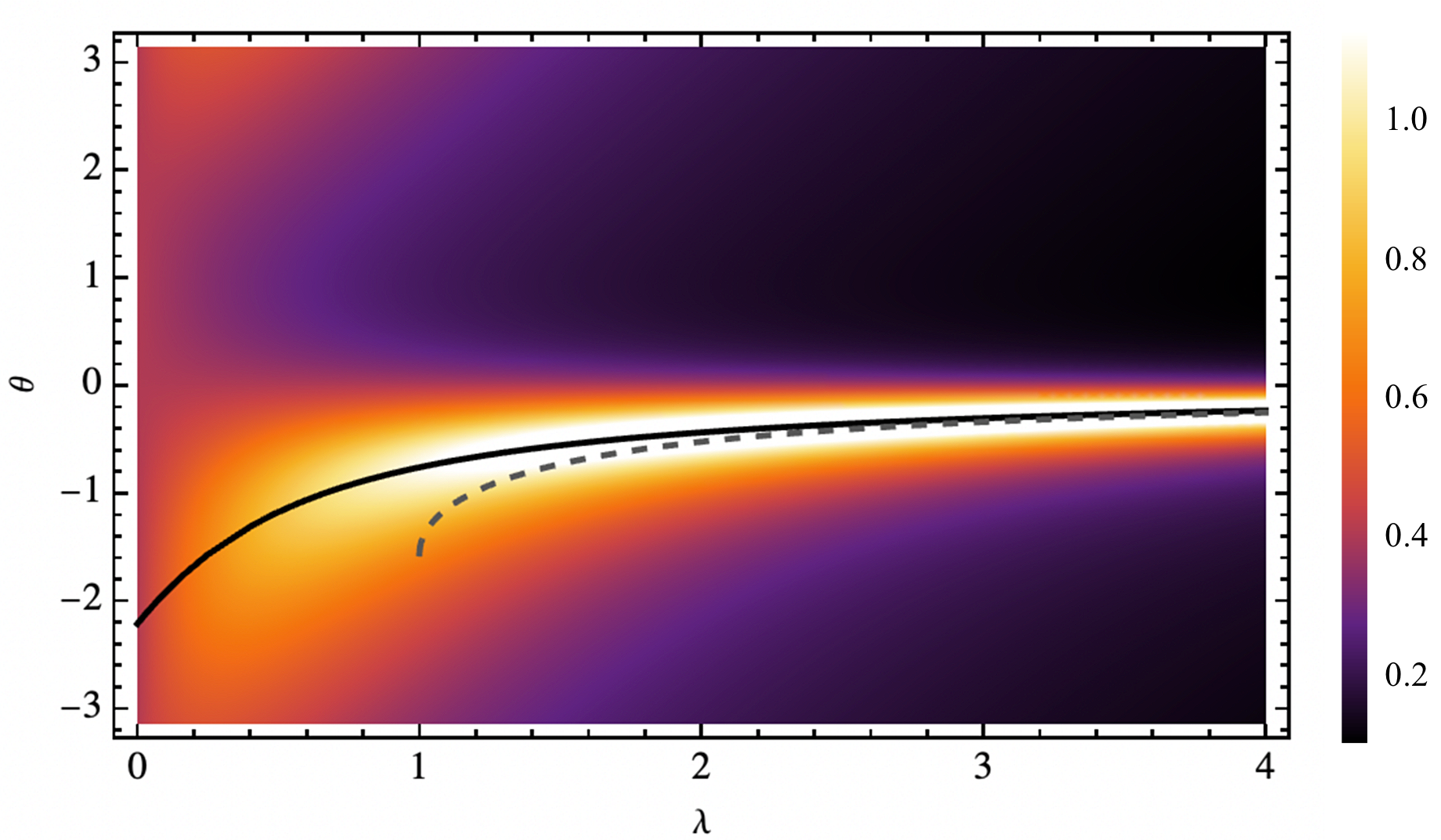}
    \caption{Effective diffusion coefficient as a function of tumbling rate $\lambda$ and tumbling angle $\theta$. {  The colorbar indicates the value of the effective diffusivity measured relative to the diffusivity $v_0^2/D_r$ of a normal active Brownian particle.} The black solid line corresponds to the optimal tumbling angle for fixed rate  [Eq. (\ref{eq:opt_th})]. The dashed line shows the tumbling angle that makes the effective chirality $\omega_\text{eff}$ vanish, e.g. $\tilde \theta = \sin^{-1}(-\omega/\lambda)$, valid only when the argument is greater than unity. Parameters are set to $D_r = 1/2, v_0 = 1, \omega = 1$.}
    \label{fig:deff_asym}
\end{figure}

\begin{equation}\label{eq:opt_th}
    \theta_* = -2\tan^{-1}\left(\frac{\omega}{D_r + 2\lambda}\right)
\end{equation}

Fig. \ref{fig:deff_theta} shows the effective diffusion coefficient as a function of $(\omega, \theta)$ together with the optimal tumbling angle $\theta_*$ for fixed $\omega$ in dashed lines for various values of the tumbling rate. We see that, generally, a positive chirality is associated with a negative optimal tumbling angle. This has the interpretation that the tumbles try to counteract the chirality, making the particle trajectories as linear as possible. However, this intuitive picture only holds in the limit of frequent tumbling angles and low angular noise. 

To better see this, consider the tumbling angle for which $\omega_\text{eff}=0$. This is simply $\tilde \theta = \sin^{-1}(-\omega/\lambda)$, which in general is not equal to the optimal tumbling angle we have derived above. However, this angle has a Taylor series which coincides with $\theta_*$ at first order in $\lambda^{-1}$. Indeed, expanding the $\sin^{-1}(x) = x+...$ we have at large rates $\tilde \theta \approx -\omega/\lambda$. Similarly, Eq. (\ref{eq:opt_th}) to first order takes the form $\theta_* \approx - \frac{2 \omega}{D_r +2\lambda}$. {  When $D_r \ll 2 \lambda$, we can further approximate $\theta_* \approx - \frac{ \omega}{ \lambda}$. Figure \ref{fig:deff_asym} shows the diffusivity as a function of $(\theta,\lambda)$, together with the optimal tumbling angle as well as the tumbling angle that eliminates the effective chirality. In the regime of large tumbling rates, we can use $\theta_* \approx - \frac{ \omega}{ \lambda}$ and expand $D_\text{eff}$ to leading order, in which case we find $D_\text{eff} \approx v_0^2/D_r$ which is the diffusivity of a linear active Brownian particle. Hence, asymmetric tumbles can counteract the effect of chirality and give rise to a diffusivity that at most can take the value of a normal active Brownian particle.}

{ \color{black}
\section{Dynamical realization of reorientations}
Finally, we present a scenario where the distribution of reorientation angles $b(\theta)$ is realized dynamically rather than assumed \textit{ad hoc}. We consider a model for intermittent active motion, where the particle stochastically switches between an active mobile phase, and an immobile phase. During the immobile phase, no translational motion is generated, while the particle's orientation follows a dynamics similar to that of a chiral active Brownian particle (see Fig. (\ref{fig:intermittent})). While such models for intermittent motion has been studied for decades \cite{singwi1960diffusive,weiss1973diffusion}, recent years has seen a renewed interest in the topic \cite{datta2024random,santra2024dynamics,doerries2022rate,doerries2022apparent,doerries2023emergent,peruani2023active,saragosti2012modeling,thiel2012anomalous,detcheverry2017generalized}.

The immobile phases may be seen as a finite-time tumble event \cite{saragosti2012modeling,datta2024random}, where the distribution of tumbling angles depend on the duration of the immobile phase. In particular, if a particle enters one of the immobile phases at time $t$ and leaves it at a time $t =t' +\tau$, we will assume the duration of stops to be drawn from a distribution with density $\psi_S(\tau)$. The particle's orientation will in this time evolve according to
\begin{equation}
    \phi(t) =\phi(t') +  \theta(\tau)
\end{equation}
where the increment $\theta(\tau)$ is independent of the absolute time and only depends on the duration of the stop. The explicit dynamics is that of a chiral active Brownian particle with chirality $\omega_S$ and rotational diffusion coefficient $D_S$, such that the increments are distributed according to
\begin{equation}\label{eq:dyn}
    b(\theta) = \int_0^\infty d \tau \psi_S(\tau) \frac{\exp\left(-\frac{(\theta-\omega_S \tau)^2}{4 D_S \tau}\right)}{\sqrt{4 \pi D_S \tau}}
\end{equation}
Here it is assumed that the dynamics in the immobile phase in general may be different from that of the mobile phase.

\begin{figure}
    \centering
    \includegraphics[width=8.5cm]{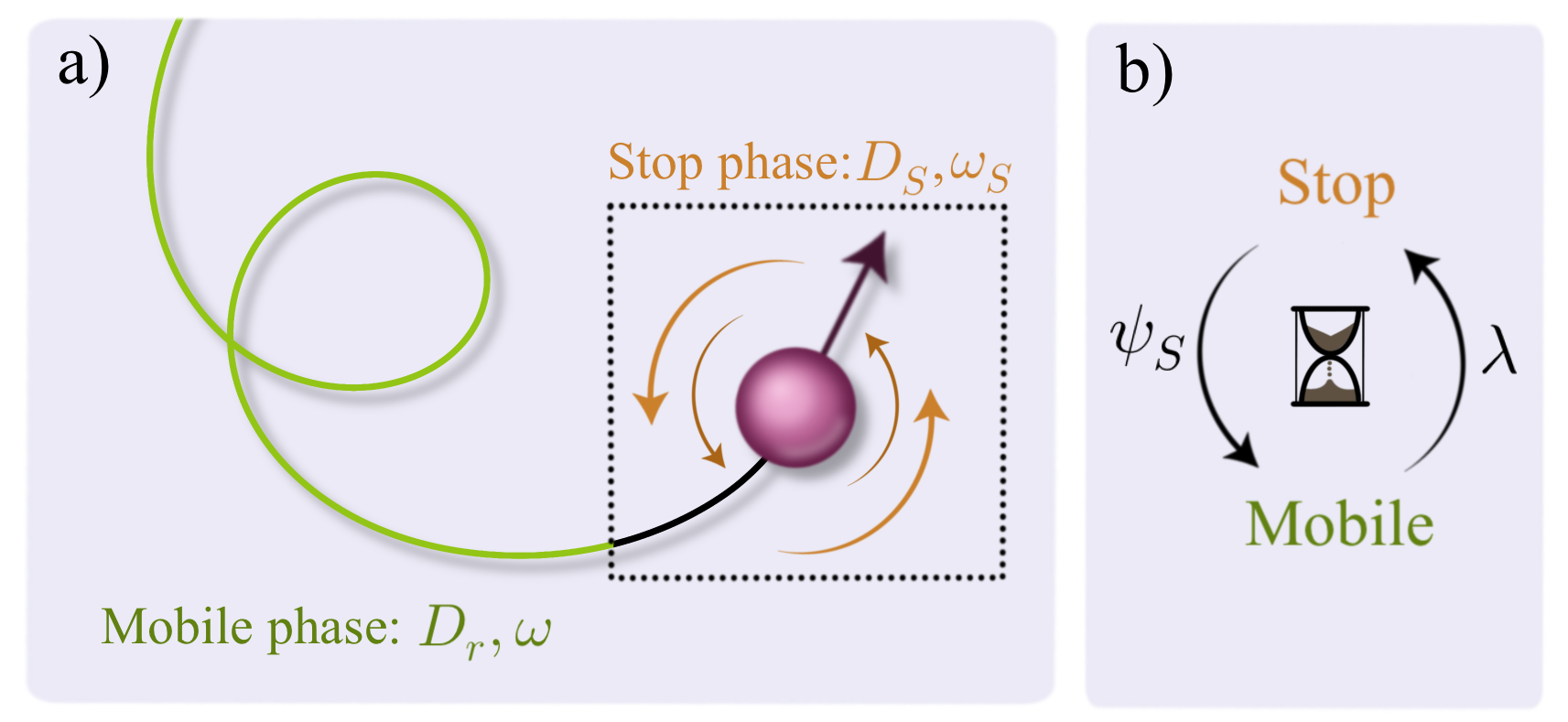}
    \caption{{ \color{black} a) Intermittent active motion consists of mobile phases and immobile ("stop") phases. During the two, the angular dynamics of the particle may in general be different. b) The particle transitions from a mobile phase to a stop phase with rate $\lambda$. The stop phases have durations drawn from $\psi_S(\tau)$.}}
    \label{fig:intermittent}
\end{figure}

Since the tumbles in this scenario have a finite duration, we must also rescale the mean squared displacement accordingly, since Eq. (\ref{eq:deff}) assumes instantaneous tumbles. To proceed, it will be beneficial to temporarily work in an ensemble of fixed number of immobile phases. A similar line of reasoning, whereby turning to a fixed-$n$ ensemble simplified calculations, has been used both in stochastic resetting and run and tumble processes in the past \cite{mori2021condensation,mori2020universal,hartmann2020convex,singh2022mean,olsen2023thermodynamic,mori2023entropy,olsen2024thermodynamic}. Indeed, in an ensemble of fixed number $n$ of immobile phases, or tumbles, the mean squared displacement asymptotically behaves as 
\begin{equation}
    \langle \bm{x}^2(n) \rangle_\text{inst.} \approx 2 D_\text{eff}^{(\text{inst.})} t_n =  2 D_\text{eff}^{(\text{inst.})} \langle \tau_m \rangle n,
\end{equation}
where we used that $n$ mobile phases take a time $ \langle \tau_m \rangle n $ when the immobile phases are instant. Here, each mobile phase has mean duration $\langle \tau_m \rangle  = 1/\lambda$. However, since no motion is generated in the immobile phases by definition, this fixed-$n$ mean squared displacement (with fixed reorientation angles) must be the same even if the immobile phases had a duration $\tau_S$. In this case, however, the mean observation time is given by $t_n = n(\langle \tau_S\rangle + \langle \tau_\text{m}\rangle)$. This leads to a rescaled mean squared displacement when the immobile phases have finite duration
\begin{equation}
    \langle \bm{x}^2(n) \rangle =  2\frac{\langle \tau_\text{m}\rangle  D_\text{eff}^{(\text{inst.})}}{\langle \tau_S\rangle+\langle \tau_\text{m}\rangle} t_n
\end{equation}
Hence for finite-time tumbles, we expect the effective diffusion coefficient to be simply rescaled as 
\begin{equation} \label{eq:deff_ft}
   D_\text{eff} =  \frac{\langle \tau_\text{m}\rangle  }{\langle \tau_S\rangle+\langle \tau_\text{m}\rangle} D_\text{eff}^{(\text{inst.})}.
\end{equation}

\begin{figure}
    \centering
    \includegraphics[width=8.6cm]{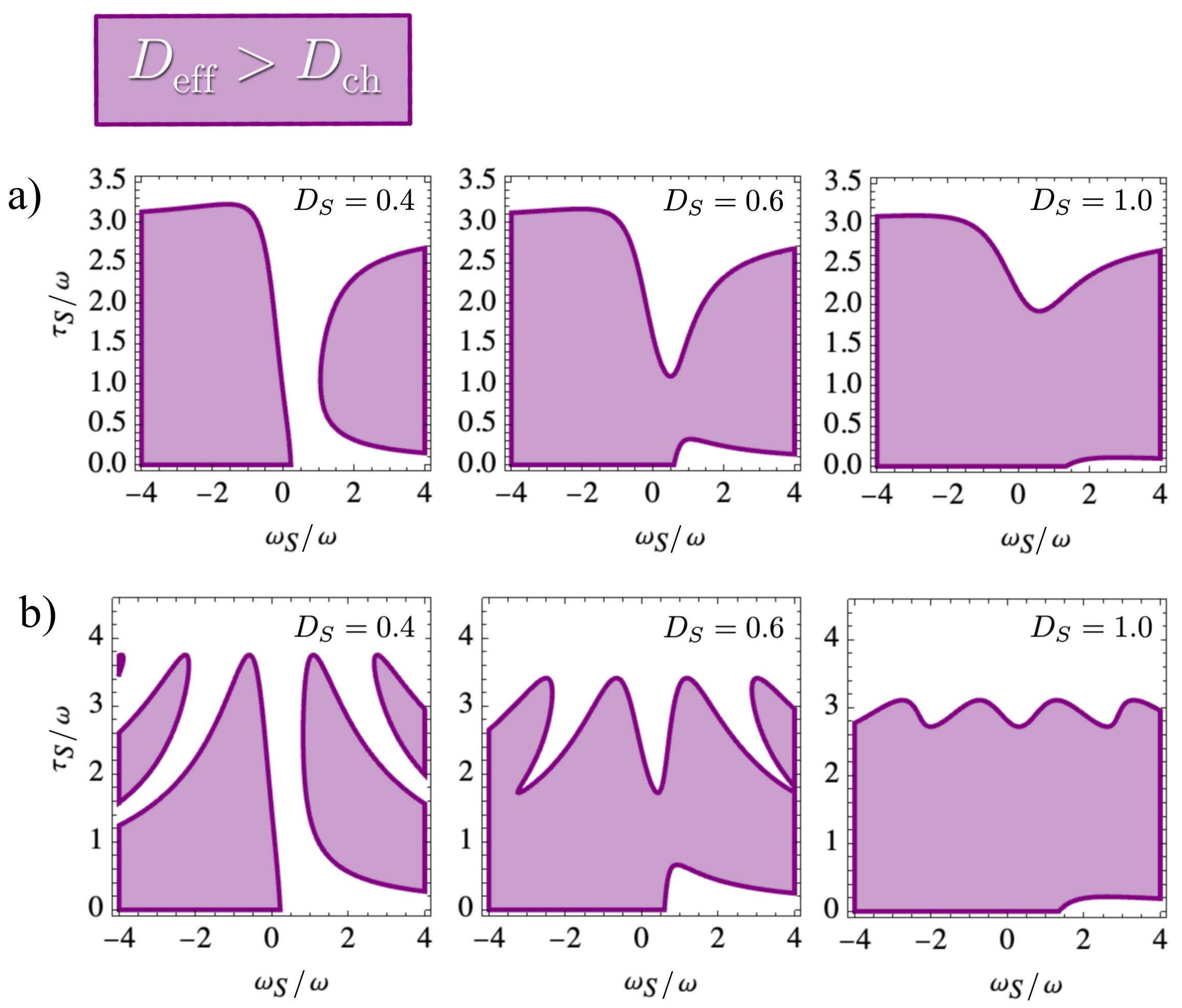}
    \caption{{ \color{black} Phase diagrams showing where diffusion is enhanced due to intermittent reorientations. a) Exponential density $\psi_S(\tau) = \exp(-\tau/\tau_S)/\tau_S$, for three values of $D_S$. b) Dirac delta density $\psi_S(\tau) = \delta(\tau -\tau_S)$, for three values of $D_S$. Other parameters are set to $\omega = 1,D_r = 1/4, \langle \tau_m\rangle = \lambda^{-1} = 5$.}}
    \label{fig:intermittent2}
\end{figure}

We see that in order to calculate the effective transport at late times for dynamically realized reorientations, Eq. (\ref{eq:deff_ft}), we need only the diffusion coefficient for instantaneous reorientations  $D_\text{eff}^{(\text{inst.})}$, given by Eq. (\ref{eq:deff}), together with the distribution of reorientation angles Eq. (\ref{eq:dyn}). The effective persistence time and effective chirality characterizing the reorientations from Eq. (\ref{eq:dyn}) can in this case be shown to take the form 
\begin{align}
    \tau_\text{eff}^{-1} &= D_r +  \lambda - \lambda \int_0^\infty d\tau \psi_S(\tau) e^{- D_S \tau}\cos(\omega_S \tau) \nonumber\\
    & = D_r +   \lambda - \lambda \Re[ \tilde{\psi}_S (D_S - i \omega_S)]\\
    \omega_\text{eff} &= \omega + \lambda \int_0^\infty d\tau \psi_S(\tau) e^{- D_S \tau}\sin(\omega_S \tau) \nonumber \\
    & = \omega +   \lambda \Im[ \tilde{\psi}_S (D_S - i \omega_S)]
\end{align}
where $\tilde{\psi}_S$ is the Laplace transform of the density of stopping durations. In the absence of chirality in either of the two phases, i.e. $\omega =\omega_S = 0$, a complementary approach based on the velocity-correlation functions of intermittent active particles was recently studied in Ref. \cite{datta2024random}. 

Figure (\ref{fig:intermittent2}) shows phase diagrams over the chirality and duration of the immobile phases $(\omega_S,\tau_S)$, where shaded regions indicate diffusion enhancement, i.e. $D_\text{eff} \geq D_\text{ch}$. Figure (\ref{fig:intermittent2}a) considers exponentially distributed immobile durations, with fixed mean $\tau_S$. Figure (\ref{fig:intermittent2}b) considers Dirac delta distributed immobile durations, with duration $\tau_S$. Common to all cases is the fact that stops have to be sufficiently short but diffusion to be enhanced. Furthermore, when the chirality in the stop phase has the same sign as in the mobile phase, we see that the durations of the stops also have to be sufficiently long. We also note that in the case of deterministic stopping times, Fig.  (\ref{fig:intermittent2}b), there is a strong re-entrant behavior as $\tau_S$ is varied for small values of $D_S$. This comes from the fact that an optimal reorientation angle can be achieved by performing an additional complete rotation, which is possible only if the noise is sufficiently small.

}

\section{Discussion}
Dynamics of active particles under the simultaneous influence of chirality and tumbling dynamics has been considered. We have derived exact expressions for the angular probability density function, as well as for the corresponding correlation function. We found that while normally rotational diffusion determines the exponential decay rate of the correlations while chirality sets the oscillatory parts, the presence of tumbles couple these two effects. Effective transport properties have been investigated, and various tumbling strategies that maximize the effective diffusion coefficient have been identified. 

For symmetric tumble distributions, we have shown that the diffusion coefficient can always be optimized to gives rise to enhanced diffusion relatively to the pure chiral case. The value of the diffusion coefficient at optimality is universal, and is independent of the detail of the tumbles. 

For asymmetric tumbles, the tumbles can be tuned to counteract the effect of the underlying chirality, giving rise to trajectories which are more rectilinear. At fixed tumble rate, optimal tumbling angles has been identified. 

{ \color{black}
Finally, we have presented a model where the reorientations are realized dynamically. The particles move intermittently, and while in immobile phases with no translational motion, reorients according to Langevin dynamics with rotational diffusion coefficient $D_S$ and chirality $\omega_S$. Parameter space regions for which diffusion can be enhanced has been discussed. 
}

In the future, it would be interesting to explore further hybrid active dynamics, or even include more than two dynamical modes as presented here. { \color{black} It would also be of interest to see whether intermittent motion in three-dimensional helical motion could lead to similar enhancements of diffusivity \cite{jennings1901significance,friedrich2007chemotaxis,wittkowski2012self,goldstein2015green,kirkegaard2016motility}.} For navigation problems where motion in a specific direction is desirable, such as in chemotaxis, it would be interesting to study the optimal reorientations that enhance the effective drift for chiral particles, not only effective diffusion \cite{leptos2023phototaxis,mori2023optimal,baouche2024active}.  Furthermore, many real-world examples of active matter reside in heterogeneous environments where optimal navigation past obstacles or other types of disorder is a crucial task \cite{bechinger2016active,bertrand2018optimized,makarchuk2019enhanced,alonso2019transport,chepizhko2020random,olsen2021active,kurzthaler2021geometric,van2022role,saintillan2023dispersion,lohrmann2023optimal,olsen2024hyper,jin2024microbes}. Whether hybrid motility patterns can be beneficial in this case should also be explored in the future.

\begin{acknowledgements}
The authors acknowledge support by the Deutsche Forschungsgemeinschaft (DFG) within the project LO 418/29-1. 
\end{acknowledgements}


%

\end{document}